\documentclass[aps,pra,twocolumn,showpacs]{revtex4}

\usepackage{graphicx} 
\usepackage{bm} 
\usepackage{epsfig}
\usepackage{amsmath}
\usepackage{amssymb}
\usepackage{hyperref}

\begin{document}

\newcommand{\be}{\begin{equation}}
\newcommand{\ee}{\end{equation}}
\newcommand{\bea}{\begin{eqnarray}}
\newcommand{\eea}{\end{eqnarray}}

\title{Restoring Heisenberg limit via collective non-Markovian dephasing}
\author{D. Mogilevtsev$^{1}$, E. Garusov$^{1}$, M. V. Korolkov$^{1}$, V. N. Shatokhin$^2$, and S. B. Cavalcanti$^{3}$}
\affiliation{$^{1}$B.I. Stepanov Institute of Physics, National Academy of Sciences of Belarus, 220072 Minsk, Belarus \\
$^2$Physikalisches Institut, Albert-Ludwigs-Universit\"at Freiburg,
Hermann-Herder-Stra\ss{}e 3, D-79104 Freiburg, Germany\\
$^{3}$Instituto de F\'{\i}sica, UFAL, Cidade Universit\'{a}ria, 57072-970,
Macei\'{o}-AL, Brazil}

\begin{abstract}
In this work an exactly solvable model of
$N$ two-level systems interacting with a single
bosonic dephasing reservoir is considered to unravel the role
played by collective non-Markovian dephasing. We show that phase estimation with entangled states for this model can exceed the standard quantum limit and demonstrate Heisenberg scaling with the number of atoms for an arbitrary temperature. For a certain class of reservoir densities of states decoherence can be suppressed in the limit of large number of atoms and the Heisenberg limit can be restored for arbitrary interrogation times. We identify the second class of densities when the Heisenberg scaling can be restored for any finite interrogation time. We also find the third class of densities  when the standard quantum limit can be exceeded only on the initial stage of dynamics in the Zeno-regime.
\end{abstract}


\maketitle

\section{Introduction}

It has been already known for quite a long time that by performing measurement on an entangled set of systems one can enhance estimation precision of system's parameters in comparison with the measurements on untangled systems \cite{caves,wineland}. Using the Ramsey interferometry scheme implemented in atomic clocks, to measure completely entangled states of $N$ atoms, one may improve a phase estimation error scaling, from $N^{-1/2}$ (which constitutes the Standard Quantum Limit, or SQL) to  $N^{-1}$, which constitutes the Heisenberg limit. However, dephasing noise can completely obliterate  this advantage. Individual Markovian decoherence of each atom reduces error scaling back to SQL \cite{huelga1,dobr1}. After the appearance of this seminal result a considerable effort has been put forward to clarify an influence of dephasing on the error scaling (see, for example, the review \cite{toth}). It was shown that individual non-Markovian dynamics can lead to the general scaling $N^{-3/4}$ for small times when the Zeno-regime holds (i.e., when the decay rate is proportional to the time) \cite{huelga2,mats,maci}. It was also recognized that the density of states of the dephasing reservoir may deeply affect error scaling, both  in the Zeno regime and for longer times \cite{mash}. Also, collective character of dephasing is able to influence the scaling. Collective dephasing may restore the Heisenberg scaling through the generation of  decoherence-free subspaces  \cite{jeske}. On the other hand, collective effects leading to spatial correlations in dephasing can be strongly disadvantageous for phase estimation with the maximally entangled states exhibiting the Heisenberg scaling for noiseless estimation \cite{dorner,jeske}.

However, for collective non-Markovian dephasing it is very important to provide an accurate description of an interplay between classical interference of the reservoir systems arising due to spatial arrangement of atoms, quantum correlations of the collective atomic state and reservoir-atomic correlations created by the atom-reservoir interaction. Non-Markovian dephasing generates long-living correlations between atoms and reservoir leading to incomplete coherence loss even for rather large temperature of the reservoir \cite{kuhn}, to preservation \cite{doll}, and to appearance of quantum correlations between initially uncorrelated atoms \cite{ros,krzywda}. For such cases one must implement for analysis not an approximate model, but an exact solution revealing dynamics of both the atoms and the reservoir.

Using an exact solution for an ensemble of atoms interacting with the same bosonic dephasing reservoir, we show that for a class of super-Ohmic reservoir densities of states interference between reservoir bosonic modes arising due to spatial arrangements of atoms can suppress dephasing of the initial entangled state of the atomic ensemble in the limit of large number of atoms for any measurement time and temperature, and restore the Heisenberg limit. It is already known that such quantum-classical
interference phenomena lead to a number of curious effects such as: precise directional emission of an absorbed photon by an atomic ensemble \cite{scully2006}, superdirectional  intensity distributions  \cite{agarwal2011},  non-reciprocity  \cite{boag2013} and possibility to generate directional entangled state by quantum antennas \cite{we2018}. The ability to partially preserve coherence
and quantum correlations by non-Markovian collective dephasing for a spatially distributed atomic system in
a Werner state has been recently demonstrated \cite{doll}.

We also specify a class of densities for which the Heisenberg limit can be restored for a finite interrogation time. Curiously, for such densities in increasing of the number of atoms, the Zeno regime can be eventually reached for any finite interrogation time. Also, even for densities for which the Heisenberg limit is not attainable, an advantage over SQL can be reached in the Zeno regime.

The outline of the paper is as follows. Section II describes the exactly solvable model of en ensemble of two-level atoms interacting with the common bosonic dephasing reservoir and presents the exact solution. Notably, even for large reservoir temperatures, the coherence dynamics can be pronouncedly non-Markovian. In  Section III, the decoherence suppression by classical interference effects is discussed. The dependence of the coherence on the number of atoms in the entangled atomic cloud initially in the Greenberg-Horne-Zeilinger (GHZ) state is discussed. Existence of the thresholds  for common power-low densities is established. Finally, Section IV discusses phase estimate using the
Ramsey interferometry scheme and the restoration of Heisenberg limit by the effects of collective non-Markovian decoherence.

\section{Dephasing model}

{To show how the collective character of non-Markovian dephasing can restore the Heisenberg limit}, we set out by writing the Hamiltonian representing the interaction
between $N$ identical qubits and a single common
dephasing reservoir of bosonic modes:
\bea
\nonumber
H=\hbar\sum\limits_{j=1}^N\omega|+_j\rangle\langle+_j|+\hbar\sum\limits_{\forall {\vec k}}w_{\vec k}a_{{\vec k}}^{\dagger}a_{{\vec k}}+\\
\hbar\sum\limits_{j=1}^N\sum\limits_{\forall {\vec k}}|+_j\rangle\langle+_j|(g_{{\vec k}}e^{-i{\vec k}{\vec r}_j}a_{{\vec k}}+h.c.),
\label{ham1}
\eea
where $|\pm_j\rangle$ is upper(lower) state of $j$-th qubit situated at the position given by the vector ${\vec r}_j$; the transition frequency of the qubits is $\omega$. A bosonic mode with the wave-vector ${\vec k}$ and the frequency $w_{\vec k}$ is described by creation, $a_{{\vec k}}^{\dagger}$, and annihilation,  $a_{{\vec k}}$, operators. The coupling of the latter to the atoms is quantified by the position-independent interaction strengths $g_{\vec{k}}$. {The simple Hamiltonian (\ref{ham1}) is typical, for example, for a system of non-interacting solid-state qubits, for instance, excitons in quantum dots \cite{haken}.}

We  assume that the initial state of the atomic ensemble is not correlated with the reservoir and is described by the Greenberger-Horn-Zeilinger (GHZ) state, that is
\be
|\psi\rangle=\frac{1}{\sqrt{2}}(|\psi_0\rangle+|\psi_N\rangle), \quad |\psi_{0(N)}\rangle=\prod\limits_{j=1}^N|-(+)_j\rangle.
\label{ghz}
\ee
The initial state of the reservoir is the thermal state described by the density matrix
\be
\rho_{res}=\frac{\exp\{-Z\}}{Tr(\exp\{-Z\})}, \quad
Z= \beta \sum\limits_{\forall {\vec k}}w_{\vec k}a_{{\vec k}}^{\dagger}a_{{\vec k}},
\label{thermal}
\ee
where $\beta=\hbar/k_{B}T$, and $T$ is the temperature of the reservoir.

From the Hamiltonian (\ref{ham1}), it is simple to calculate the coherence of the reservoir-averaged density matrix \cite{breuer,krzywda}:
\bea
\nonumber
\rho_{0,N}(t)=Tr\left(\langle\psi_0|e^{-iHt/\hbar}\rho_{res}\otimes|\psi\rangle\langle\psi|e^{iHt/\hbar}|\psi_N\rangle\right)=\\
\exp(iN\omega t+i\Delta_N(t)-\gamma_{N}(T,t))\rho_{0,N}(0), \quad
\label{coh}
\eea
where the dephasing factor is manifestly non-Markovian \cite{breuer}, given by
\be
\gamma_{N}(T,t)=\sum\limits_{\forall {\vec k}}\frac{1-\cos(w_{\vec k}t)}{w_{\vec k}^2}
\coth\left(\frac{\beta w_{\vec k}}{2}\right)|g_{{\vec k}}G_N({\vec k})|^2,
\label{gam1}
\ee
and the function describing the effect of classical correlations of the reservoir modes is
\be
G_N({\vec k})=\sum\limits_{j=1}^Ne^{-i{\vec k}{\vec r}_j}.
\label{g}
\ee
For the initial GHZ state of the multi-qubit system, the frequency shift induced by the reservoir is temperature-independent:
\be
\Delta_N(t)=-\sum\limits_{\forall{\vec k}}\frac{|g_{\vec k}G_N({\vec k})|^2}{w_{\vec k}}\left(t-\frac{\sin(w_{\vec k}t)}{w_{\vec k}}\right).
\label{shift}
\ee

\section{Suppressing decoherence}

Here, we demonstrate that the interference of the phase terms in the r.h.s. of Eq. (\ref{g}) due to the random distribution of atoms, is able to suppress decoherence. The key observation here is that the  function $G_N({\vec k})$ can be quite close
to the delta-function from ${\vec k}$ for a larger number of randomly placed atoms \cite{scully2006}.  So, for such an ensemble of atoms, one can intuitively expect diminishing the dephasing factor $\gamma_{N}(T,t)$ with increasing $N$, for fixed $T$ and $t$. For quantifying this intuition, let us assume a simple isotropic 3D Gaussian density of the atomic cloud
\be
\varrho({\vec r})=\frac{1}{(2\pi \sigma^2)^{3/2}}\exp\left(-\frac{|{\vec r}|^2}{2\sigma^2N^{2/3}}\right),
\label{gauss}
\ee
satisfying $\int d^3r \varrho({\vec r})=N$, with $\sigma$ defining the root mean square distance between atoms. Thus, taking
\be
G_N({\vec k})\approx \int d^3r \varrho({\vec r})e^{-i{\vec k}{\vec r}},
\label{cont1}
\ee
and assuming the isotropic linear dispersion relation for the reservoir, $w_{k}=c|{\vec k}|$, we arrive at the following result for the dephasing factor:
\bea
\nonumber
\gamma_{N}(T,t)={N^2}\int\limits_{0}^{+\infty}dw J(w)(1-\cos(wt))\times \\
\coth\left(\frac{\beta w}{2}\right)\exp\left(-\frac{w^2\sigma^2}{c^2}N^{2/3}\right),
\label{gam2}
\eea
where for a typical 3D isotropic power-law density (see, for example, \cite{ram}),
\be
J(w)=\alpha_sw^s\exp\left(-\frac{w^2}{w_s^2}\right),
\label{density}
\ee
and $w_s$ is the cut-off frequency.{Notice that the quantity  (\ref{density}) is defined by the spectral density of the reservoir modes and individual qubit parameters. It does not depend on the distribution of qubits.}

\subsection{Zero temperature}
Let us first consider  zero-temperature case, $T=0$.
Under the condition that the cut-off frequency $w_s\gg c/\sigma N^{1/3}$, for $s>-1$
one arrives at
\bea
\nonumber
\gamma_{N}(0,t)\approx\frac{\alpha_s}{2}N^{2-(s+1)/3}{\bar w}^{s+1}\Gamma\left(\frac{s+1}{2}\right)!\times \\
\left(1-\exp\left(-\frac{{\bar w}^2t^2}{4N^{2/3}}\right)M\left(-s/2,1/2,\frac{{\bar w}^2t^2}{4N^{2/3}}\right)\right),
\label{solution0}
\eea
where ${\bar w}={c}/{\sigma}$, $\Gamma(x)$ is the Gamma function, and $M(x,y,z)$ is the confluent hypergeometric (Kummer's) function, with the following asymptotic (for $|z|\rightarrow \infty$) behaviour \cite{nist}
\be
M(x,y,z)\Bigl|_{|z|\rightarrow+\infty}\approx \frac{\Gamma(y)}{\Gamma(x)}e^zz^{x-y}+\frac{\Gamma(y)}{\Gamma(y-x)!(-z)^x}.
\label{longtime}
\ee

Now, one may derive a number of important conclusions by inspecting Eqs. (\ref{solution0},\ref{longtime}). In the long-time limit the dephasing factor always tends to the constant non-zero value for densities \eqref{density} with $s>-1$. This value is
\[\gamma_{N}(0,+\infty)\approx \frac{\alpha_s}{2}N^{2-(s+1)/3}{\bar w}^{s+1}\Gamma\left(\frac{s+1}{2}\right).\]
For a single qubit, the solution is given by Eq.(\ref{solution0}) with ${\bar w}$ replaced with $w_s$, and $N=1$. So, we have a similar expression for the asymptotic dephasing factor:
\[\gamma_{1}(0,+\infty)\approx \frac{\alpha_s}{2}{w}_s^{s+1}\Gamma\left(\frac{s+1}{2}\right).\]
Obviously, for a fixed number of qubits, $N$, and $s> -1$, it is always possible to achieve a lower asymptotic dephasing factor than for a single qubit by adjusting their spatial distribution (varying $\sigma$), as long as approximations of non-interacting qubits and common dephasing reservoirs hold. But the most interesting thing for us is the behaviour of the dephasing factor with increasing the number $N$ of atoms. Obviously, for $s>5$ the value of $\gamma_{N}(0,+\infty)$  tends to zero with increasing $N$. Thus, for such super-Ohmic densities of reservoir states, dephasing can be completely suppressed for sufficiently large number of atoms in the ensemble being in the GHZ state (which is not captured by the models considered in Refs. \cite{dorner,jeske}).

So, classical interference in the  process of collective non-Markovain decoherence can preserve entanglement. Notice that the discussion above can be easily re-iterated for the 1D case. In the latter case,  dephasing is suppressed with increasing $N$ for $s>3$.

{It is worth noting that densities with the  values of $s$ required for suppressing decoherence
can be reached in practice. For zero temperature it can be achieved for spatially
localized qubits and the piezoelectric interaction with bulk acoustic
phonons for the Zeno-regime ($s = 3$),  and for the deformation
potential coupling with bulk acoustic phonons for the regime where the
all-time suppression is possible $s = 5$ \cite{haken}.}

\begin{figure}
  \includegraphics[width=\linewidth]{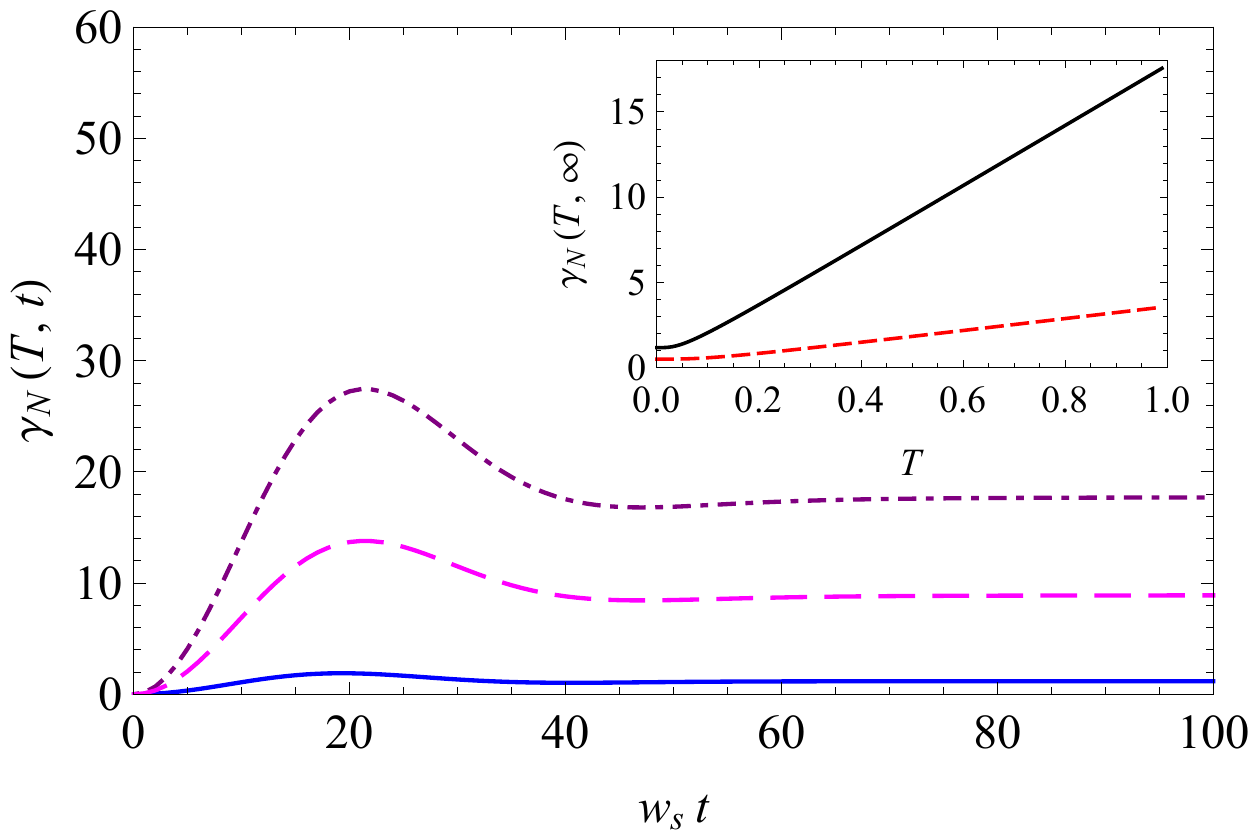}
\caption{(color online) The dephasing factor $\gamma_N(T,t)$ as a function of $w_st$, [see Eq. (\ref{gam2})], for the density (\ref{density}) with $s=4$, for different temperatures $T$ and for $N=10^3$, ${\bar w}=w_4$, and $\alpha_4w_4^5\Gamma(5/2)/2=0.12$. Solid, dashed and dash-dotted  lines correspond to $T=0,0.5,1$ (in units of $\hbar{\bar w}/k_{B}$), respectively. The inset shows the dependence of the stationary value of the dephasing factor  on  $T$, for $N=10^2$ (dashed line) and $N=10^3$ (solid line).}
\label{figure_1}
\end{figure}

\subsection{Time dynamics}

The dynamics described by Eqs. (\ref{gam2},\ref{solution0}) is also rather non-trivial. {The dephasing factor changes non-monotonically in time (see Fig. \ref{figure_1}). A similar behavior was found, for example, in Ref.\cite{kuhn}}. The character of dynamics strongly depends on the density of states (i.e., on $s$). As follows from Eq.(\ref{longtime}), if $s$ is not an even number, the dephasing factor approaches the stationary value polynomially,
\[\gamma_{N}(0,\infty)-\gamma_{N}(0,t\rightarrow\infty)\propto ({\bar w t})^{-s-1}.
\]
For even  $s$, the coherence decays as
\[
\gamma_{N}(0,\infty)-\gamma_{N}(0,t\rightarrow\infty)\propto
({\bar w}t)^s\exp\left(-\frac{{\bar w}^2t^2}{4N^{2/3}}\right).\]
It should be noted that approaching the stationary value for collective decoherence can be much slower that for the single-atom decoherence even for the case when the stationary value drops down with increasing $N$. Indeed, the time-scale is defined by the inverse cut-off frequency ($w_s$ for single atom case, and ${\bar w}/N^{1/3}$ for $N$-atom case). The Zeno-regime defined by the validity of the approximation $1-\cos(wt)\approx w^2t^2/2$ can be much longer for collective dephasing than for single-atom dephasing. In the Zeno-regime, we can have suppression of dephasing with increasing $N$ even in the case when the stationary value decreases with $N$, since the threshold in the Zeno-regime is $s>3$.

\subsection{Finite temperature}
\label{sec:fin_temp}
Generally, the dephasing factor tends to increase with the temperature $T$. For us, it is important that  finite temperature shifts
the values of $s$ for which the suppression of decoherence occurs with increasing $N$. Indeed, for large $N$, only small frequencies contribute to the integral (\ref{gam2}). So, for  finite temperature,
\[\coth\left(\frac{\beta w}{2}\right)\Bigr|_{w\rightarrow 0}\propto \frac{T}{w}.\]
Therefore, the decoherence suppression with increasing $N$ occurs only for $s>6$. In the long-time limit, the dephasing factor is directly proportional to the temperature. For finite temperatures, the threshold in the Zeno-regime is $s=4$. This means that in the 1D case  any super-Ohmic density allows for the suppression of the collective dephasing in the Zeno-regime.

\section{Measurement and error estimate}

Now, let us consider the influence of the discussed collective dephasing effects on phase estimation.
We take the measurement arrangement typical for Ramsey interferometry \cite{wineland}. We assume that initially an ensemble of $N$ qubits is prepared in a pure state. Thereafter, this state is allowed to evolve and decohere, as discussed in the previous section, and, finally, it is subjected to a measurement. We consider two scenarios: when $N$ qubits are prepared in uncorrelated states $(|-\rangle+|+\rangle)/{\sqrt{2}}$, and sent one by one to the measurement stage; and when they are prepared in the entangled GHZ-state (\ref{ghz}), and sent together to the measurement stage, decohering collectively on the way. Thus, for both cases the probability to find all the qubits in the ground states is  \cite{huelga1,huelga2}
\be
p_N(T,t)=\frac{1}{2}\left(1-e^{-\gamma_N(T,t)}\cos(N\phi t+\Delta_N(t))\right),
\label{prob}
\ee
where $\phi$ is the detuning between the atomic transition frequency $\omega$ and the frequency of the external oscillator;  the detuning induced by the dephasing, $\Delta_N(t)$, is given by Eq.(\ref{shift}).  As in Refs.\cite{huelga1,huelga2}, we assume the simplest binary measurement scheme described by the probabilities $p_N(T,t)$ and $1-p_N(T,t)$ for some particular value of $\phi$, and derive the error bound for this phase estimation. Such a bound can be derived in a standard way from the Fisher information
\begin{align}
F_N(T,t)&=\frac{1}{p_N(T,t)}\left[\frac{d}{d\phi}p_N(T,t)\right]^2\nonumber
\\
&+\frac{1}{1-p_N(T,t)}\left[\frac{d}{d\phi}(1-p_N(T,t))\right]^2
\label{fisher}
\end{align}
and from the Cramer-Rao inequality for any unbiased estimate of the detuning $\phi$. This inequality for $N$ uncorrelated qubits, measured one by one, reads
\be
\delta^2_1\phi\le \left[F_1\right]^{-1}=\frac{4p_1(T,t)(1-p_1(T,t))}{Nt^2\sin^2(\phi t+\Delta_1(t))}e^{2\gamma_1(T,t)},
\label{cramer1}
\ee
whereas for $N$ simultaneously measured qubits in the GHZ state, it reads
\be
\delta^2_N\phi\le \left[F_N\right]^{-1}=\frac{4p_N(T,t)(1-p_N(T,t))}{N^2t^2\sin^2(N\phi t+\Delta_N(t))}e^{2\gamma_N(T,t)},
\label{cramerN}
\ee
where $\delta^2_1\phi$ and $\delta^2_N\phi$ are variances of the estimated detunings.

\begin{figure}
  \includegraphics[width=\linewidth]{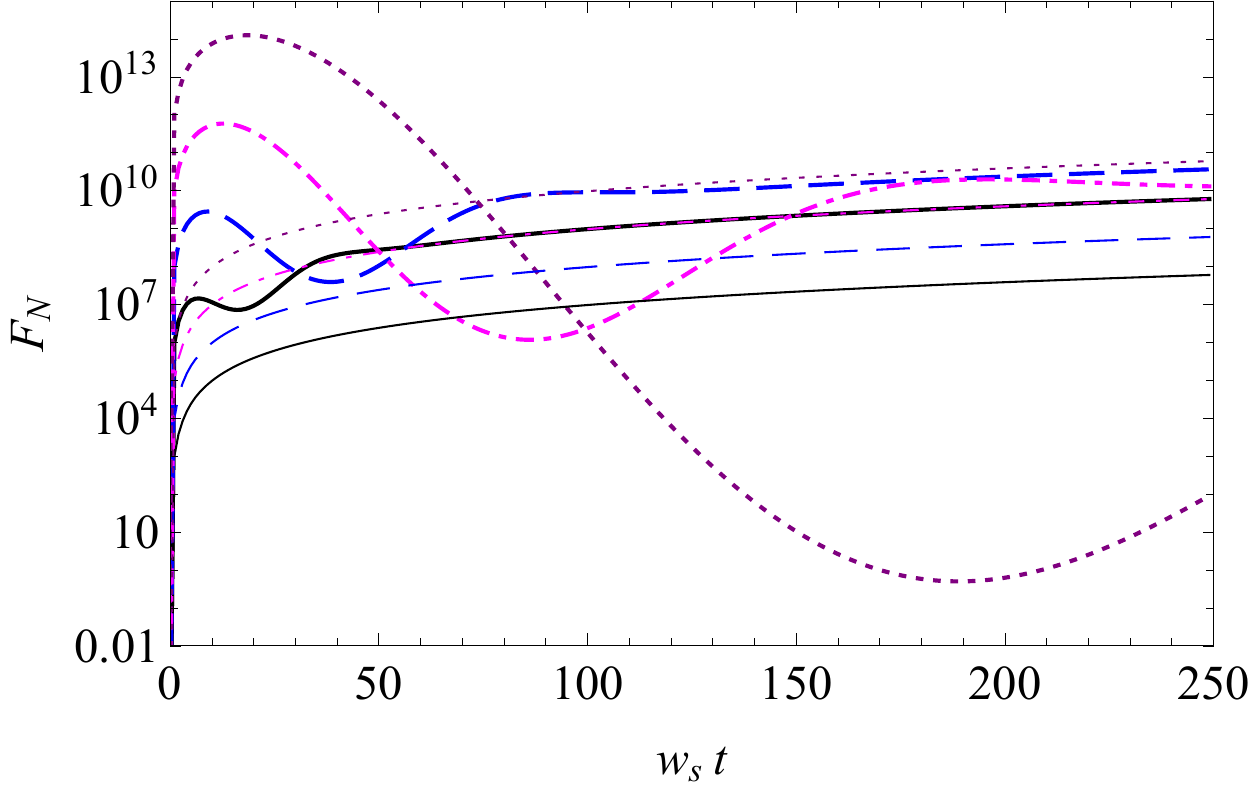}
\caption{(color online) The Fisher information $F_N(0,t)$ at the optimal detuning $\phi$ as a function of $w_st$ [see Eq. (\ref{FNF1})], for the spectral density (\ref{density}) with $s=4$, at $T=0$, ${\bar w}=w_4$ and $\alpha_4w_4^5\Gamma(5/2)/2=0.12$. Thin (thick) lines correspond to one-by-one measurements with uncorrelated initial state (to the GHZ state with $N$ simultaneously measured qubits). Solid, dashed, dash-dotted and dotted lines correspond to $N=10^3,10^4,10^5,10^6$, respectively.}
\label{figure_2}
\end{figure}

Now, let us assume that  for both cases, we have chosen the best values of the detuning $\phi$ and time $t$ that are maximizing
the Fisher information. As follows from Eqs. (\ref{cramer1},\ref{cramerN}),  for any time $t$, the best detuning is given by $\phi t+\Delta_1(t)=(2n+1)\pi/2$ for the single-qubit case and by $N\phi t+\Delta_N(t)=(2n+1)\pi/2$ for the $N$-qubit case, with $n$ an arbitrary integer number. Thus, from Eqs. (\ref{prob},\ref{cramer1},\ref{cramerN}), the Fisher information at the optimal detuning is
\begin{align}
\nonumber
F_{1}(T,t)&=Nt^2\exp\{-2\gamma_{1}(T,t)\},\\
F_{N}(T,t)&=N^2t^2\exp\{-2\gamma_{N}(T,t)\},
\label{FNF1}
\end{align}
and the best time, $t_{best}$, in the interval $(0,t_{max}]$ is given either by the solution of the equation
 \be
t\frac{d}{dt}\gamma_{1(N)}(T,t)\Bigr|_{t=t_{best}}=1,
 \label{best}
\ee
or by the maximal possible measurement time $t_{max}$.

Notice that Eq.(\ref{best}) generally gives a number of solutions corresponding to different extremums of the Fisher information. We illustrate this behavior in Fig. \ref{figure_2}, which depicts the Fisher information for the sub-threshold  density  (\ref{density}) with $s=4$ for zero temperature and different numbers of qubits. Thus, we only seek the solutions of Eq. \eqref{best} that correspond to the global maxima of the Fisher information.

\begin{figure}
  \includegraphics[width=\linewidth]{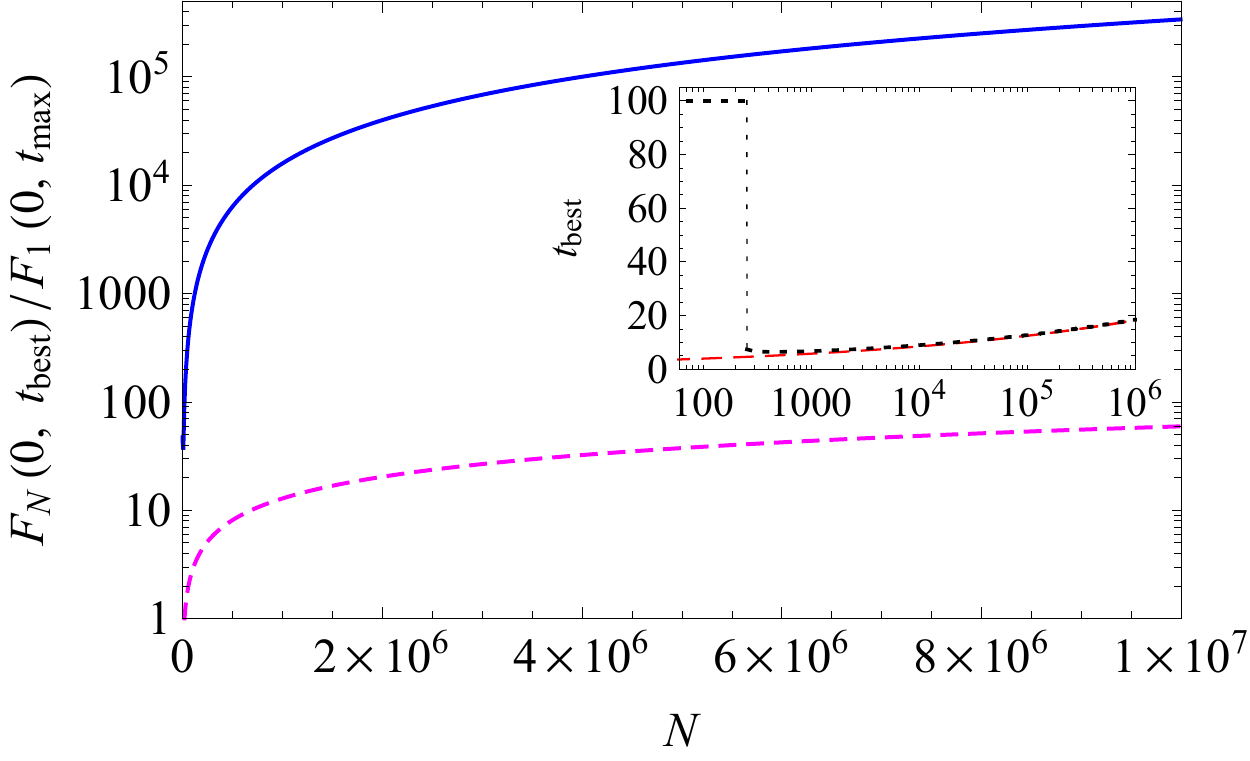}
\caption{(color online) The ratio $F_N(0,t_{best})/F_1(0,t_{max})$ versus $N$ [see Eq. \eqref{FNF1}]. The solid line corresponds to the density (\ref{density}) with $s=4$ and to ${\bar w}=w_4$, $\alpha_4w_4^5\Gamma(5/2)/2=0.12$, $t_{max}=100w_4^{-1}$. The dashed line corresponds to the density  (\ref{density}) with $s=2$ and to
 ${\bar w}=w_2$, $\alpha_2w_2^3\Gamma(3/2)/2=0.02$, $t_{max}=100w_2^{-1}$. The inset shows the optimal measurement time $t_{best}$ (in units of $w_4^{-1}$) versus $N$, for $s=4$. The
 switching from $t_{best}=100$ to $t_{best}=7.364$ occurs at $N=254$. Dotted and dashed lines show the exact numerical solution of Eq. \eqref{best} and the analytical approximation (\ref{tbest0}), respectively.}
\label{figure_3}
\end{figure}

\subsection{Restoring Heisenberg scaling for arbitrary times}

If the density of reservoir states (\ref{density}) is  above the threshold (i.e. $s=5$ for $T=0$ and $s=6$ for finite $T$), the Heisenberg scaling
is obviously restored for any moment of time, $t$, in the limit of large $N$. Indeed, for such densities, $\gamma_N(T,t)\rightarrow 0$ as $N\rightarrow\infty$, and the bound (\ref{cramerN}) scales as $N^{-2}$.

If the interrogation time $t_{max}$ significantly exceeds the time required for a single qubit to reach the stationary coherence value (this time-scale is  defined by $w_s^{-1}$), the best time for the one-by-one measurement is given by $t_{max}$. Hence, the ratio $F_N/F_1$ of the Fisher information for the collective measurement and that for the one-by-one measurement, in the limit of large $N$, is $N\exp(2\gamma_1(T,t_{max}))$. The decoherence factor increases with increasing of the reservoir temperature, such that, for finite temperatures, a measurement advantage provided by collective decoherence can be very large.

\begin{figure}
  \includegraphics[width=\linewidth]{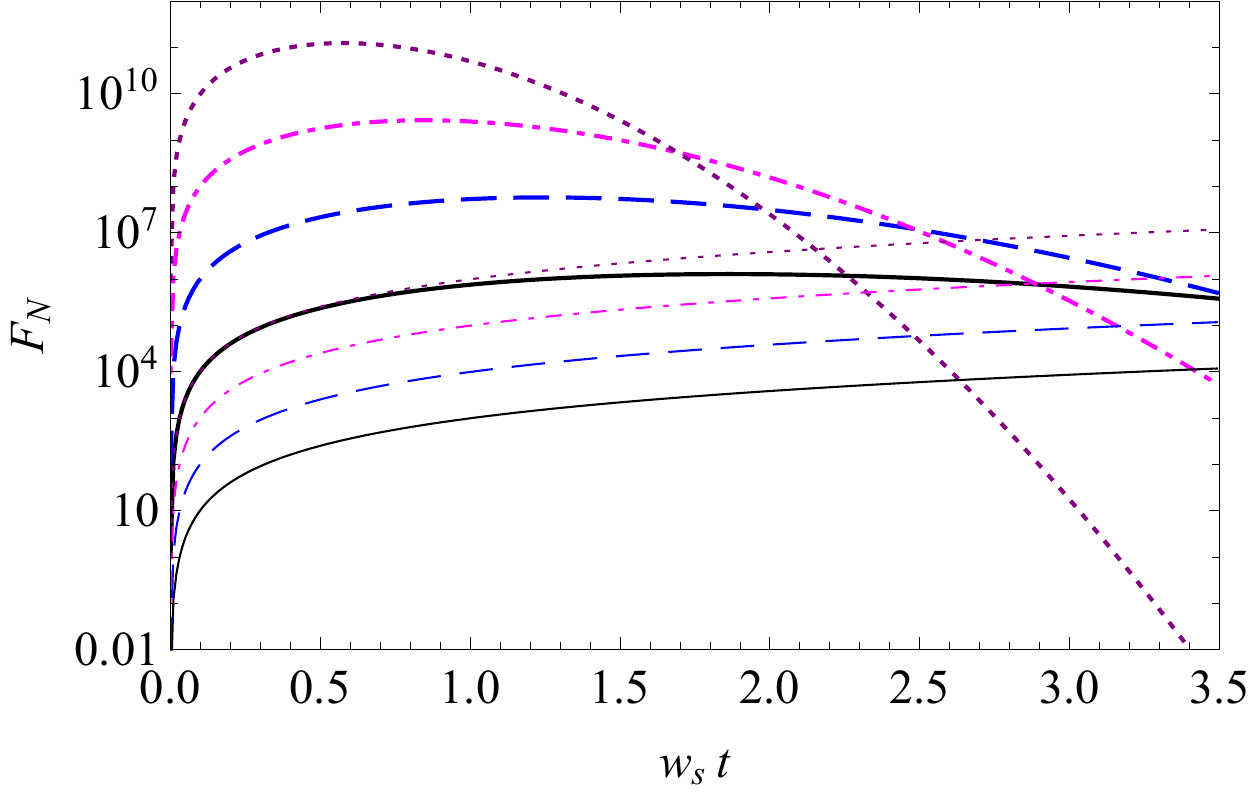}
\caption{(color online) The Fisher information $F_N(0,t)$ at the optimal detuning $\phi$ as a function of $w_st$ [see Eq. (\ref{FNF1})], for the spectral density (\ref{density}) with $s=2$, at $T=0$, ${\bar w}=w_2$ and $\alpha_2w_2^3\Gamma(3/2)/2=0.12$. Thin (thick) lines correspond to one-by-one measurements with uncorrelated initial state (to the GHZ state with $N$ simultaneously measured qubits). Solid, dashed, dash-dotted and dotted lines correspond to $N=10^3,10^4,10^5,10^6$, respectively.}
\label{figure_4}
\end{figure}

\subsection{Zeno-regime for zero temperature}

Realistically, the maximal interrogation time is limited. We already showed in the previous section that, for zero temperature in the Zeno-regime, the threshold for the density  in the 3D case corresponds to $s=3$. Thus, in the Zeno-regime above the ``Zeno threshold", the dephasing factor tends to zero for $N\rightarrow\infty$, and the Heisenberg scaling can be restored. It is curious that for any fixed maximal time, $t_{max}$, the Zeno regime would be eventually reached with increasing $N$ (this directly follows from Eqs. (\ref{gam2},\ref{solution0})). This tendency can be seen in Fig. \ref{figure_2}, where the Fisher information $F_N(0,t)$ is shown for the finite time-interval: the first maximum of the Fisher information shifts to larger times with increasing of $N$.  Another tendency can also be seen in Fig. \ref{figure_2}: for smaller $N$, the global maximum of the Fisher information is attained at $t_{best}=t_{max}$. However, with increasing $N$ the global maximum of the Fisher information is reached at $t_{best}<t_{max}$ which is determined from Eq.(\ref{best}). The switching from the high to the low value of $t_{best}$ is sharp.  This step is illustrated in the inset of Fig. \ref{figure_3}, where the time $t_{best}$ is shown for the density  (\ref{density}) with $s=4$, i.e. above the ``Zeno threshold", and for the  maximal time $t_{max}=100\omega_4^{-1}$. Figure \ref{figure_3} (solid line)  shows also the ratio  $F_N(0,t_{best})/F_1(0,t_{best})$ corresponding to the optimal measurement times $t_{best}$ for both cases (for $F_1$,  $t_{best}=t_{max}$). The value of $F_N(0,t_{best})$ is larger than that of $F_1(0,t_{max})$ for all $N$. Furthermore,  the ratio $F_N(0,t_{best})/F_1(0,t_{best})$ grows linearly for large $N$ (note the log-scale of the $y$-axis in Fig. \ref{figure_3}).

Surprisingly, even below the ``Zeno threshold" one can still get an advantage from collective dephasing. This happens because for $s>0$ the Fisher information $F_N(0,t)$ still grows quicker with time than $F_1(0,t)$ for some initial time-interval. This situation is illustrated in Fig. \ref{figure_4} for $s=2$, where $F_N(0,t)$ and $F_1(0,t)$ are plotted  in the interval $t\leq t_{max}=3.5w_2^{-1}$. Opposite to the  above ``Zeno threshold" case, the time $t_{best}$ for which the global maximum of $F_N(0,t)$ is reached now becomes smaller with increasing $N$, whereas for $F_1(0,t)$ this time is fixed at $t_{max}$. Therefore, although the ratio $F_N(0,t_{best})/F_1(0,t_{max})$ increases with $N$ and the measurement result is better than the SQL (this behaviour is illustrated in Fig. \ref{figure_3} with the dashed line),  the Heisenberg limit is not reached in this case.

From Eq. (\ref{best}), one can obtain a simple estimate for the best time in the Zeno regime. Making an ansatz $\gamma_N(T,t)\approx t^2f(T,N)$, one obtains $t_{best}\approx (2f(T,N))^{-1/2}$. Thus, for zero temperature the estimation for the best measurement time for the generic density (\ref{density}) reads
\begin{equation}
t_{best}^{(s)}\approx \left( \frac{\alpha_s}{2} N^{1-s/3}{\bar w}^{s+3}\Gamma\left(\frac{s+3}{2}\right)\right)^{-1/2}.
\label{tbest0}
\end{equation}

For example, using Eq. (\ref{best}) for $s=2$, it is possible to show that for $T=0$ and for large $N$, $t_{best}\propto N^{-1/6}$ (which is consistent with Fig. \ref{figure_3}, where $F_N(0,t_{best})/F_1(0,t_{max})\propto N^{2/3}$). The inset of Fig.3 shows the analytical approximation (\ref{tbest0}) of the best time for $s=4$. One can see that the formula (\ref{tbest0}) is quite precise for $N>10^3$.
When $s$ approaches the ``Zeno threshold", the advantage over the SQL tends to the Heisenberg scaling.
Curiously, as Eq. (\ref{tbest0}) shows, for zero temperature an advantage over the SQL can still be obtained for any $s>0$.

\begin{figure}
  \includegraphics[width=\linewidth]{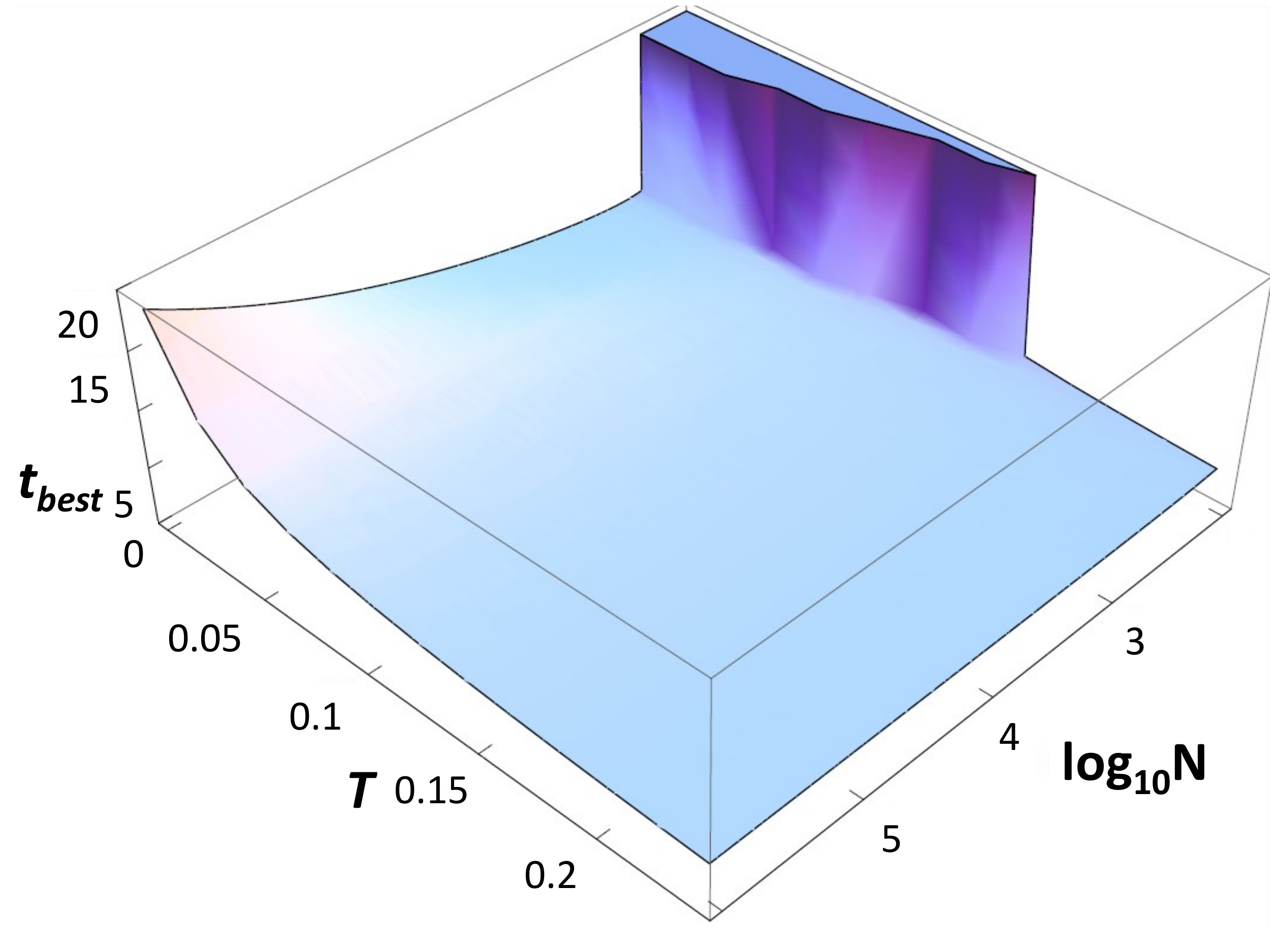}
\caption{The dependence of the best time $t_{best}$ on the temperature $T$ (in units of $\hbar\bar w/k_{B})$  and on the number of qubits $N$ (in the log-scale), for $s=4$, $t=20w_4^{-1}$, $\bar{w}=w_4$, and $\alpha_4w_4^5\Gamma(5/2)/2=0.12$.}
\label{figure_5}
\end{figure}

\subsection{Zeno-regime for finite temperature}
\label{sec:zeno_finite}
As  demonstrated in Sec. \ref{sec:fin_temp},  for finite temperature in the Zeno-regime, the threshold for the density  in the 3D case corresponds to $s=4$. So, for $4<s<6$ and for sufficiently large $N$ the Zeno regime is asymptotically attained for any finite interrogation time. Obviously, at finite temperatures the ratio  of $F_N/F_1$  increases with $N$ for $s>4$, just as it does for the corresponding zero-temperature case.
The best time  also increases with $N$. Similarly to Eq.(\ref{tbest0}), one obtains in the large $N$ limit
\begin{equation}
t_{best}^{(s)}\approx \left( \frac{\alpha_s}{2\hbar} k_BTN^{(4-s)/3}{\bar w}^{s+1}\Gamma\left(\frac{s+2}{2}\right)\right)^{-1/2},
\label{tbestT}
\end{equation}
which shows that  $t_{best}$ decreases with the temperature. Also, the Fisher information decreases with increasing $T$ (see Sec. \ref{sec:fin_temp} and Eq. \eqref{FNF1}). Therefore the value of $N$ at which the switching of the optimal measurement time happens, from the maximal  interrogation time $t_{max}$ to $t_{best} \in (0,t_{max}]$, also increases in comparison with the zero-temperature case. This can be seen in Fig. \ref{figure_5} for $t_{best}$, found from the exact numerical solution of \eqref{best} at $s=4$.  It is interesting that, despite of $s=4$ being exactly the Zeno threshold, for moderate values of $T$ and $N$, the best time still grows with $N$.

\section{Conclusions}

We have shown that collective effects in non-Markovian
dephasing do indeed restore the Heisenberg scaling in
phase estimates via a Ramsey interferometry scheme,
based on an ensemble of qubits prepared in the GHZ state.
We have demonstrated it with the help of the exactly solvable
model of an ensemble of spatially distributed qubits
interacting with a common dephasing bosonic reservoir.
In the case of the usual power-law density of states with
an exponential cut-off,
we have established the presence of two thresholds. There is
a class of super-Ohmic densities yielding the retrieval of  the
Heisenberg scaling for arbitrary times.  Then, there is a class of densities for which the Heisenberg scaling is eventually reached for any finite pre-defined interrogation time. This occurs because  with  increasing of the number of entangled qubits the Zeno regime is asymptotically attained. Below this Zeno-threshold, one can still obtain an advantage over the SQL even for sub-Ohmic densities.

\begin{acknowledgements}
D.M., E.G. and M.V.K  acknowledge support from the EU project Horizon-2020 SUPERTWIN id.686731, the National Academy of Sciences of Belarus program
"Convergence" and BRRFI grant F17Y-004. S.B.C. acknowledges the support of the Brazilian agencies, CNPq e FAPEAL.
\end{acknowledgements}

\end{document}